\documentclass[%
amsmath,
amssymb,
pra,
nourl,
]{revtex4-2}

\usepackage[dvipdfmx]{graphicx}
\usepackage{hyperref}
\usepackage{bm}
\usepackage{algorithm}
\usepackage[noend]{algpseudocode}

\newcommand{\mbfalpha}{{\bm \alpha}}
\newcommand{\mbfk}{{\bf k}}
\newcommand{\mbfx}{{\bf x}}

\newcommand{\mbfa}{{\bf a}}

\newcommand{\mbfd}{{\bf d}}
\newcommand{\mbfg}{{\bf g}}

\newcommand{\mbfn}{{\bf n}}

\newcommand{\mbfR}{{\bf R}}

\newcommand{\mbfM}{{\bf M}}

\usepackage{amsmath}
\DeclareMathOperator{\Tr}{Tr}
\begin{document}

\title{
Training Gaussian boson sampling by quantum machine learning
}


\author{Claudio Conti}%
 \email{claudio.conti@uniroma1.it}
\affiliation{%
Department of Physics, University Sapienza, P.le Aldo Moro 5, 00185 Rome, Italy
}%
\affiliation{%
  Institute for Complex Systems, National Research Council (ISC-NR),
  Via dei Taurini 19, 00185 Rome, Italy
}%
\affiliation{%
  Research Center Enrico Fermi, Via Panisperna 89a, 00184 Rome (Italy)
}%


 \homepage{https://www.complexlight.org}

\date{\today}

\begin{abstract}
We use neural networks to represent the characteristic function of many-body Gaussian states in the quantum phase space. By a pullback mechanism,  we model transformations due to unitary operators as linear layers that can be cascaded to simulate complex multi-particle processes. We use the layered neural networks for non-classical light propagation in random interferometers, and compute boson pattern probabilities by automatic differentiation. This is a viable strategy for training Gaussian boson sampling. We demonstrate that multi-particle events in Gaussian boson sampling can be optimized by a proper design and training of the neural network weights. The results are potentially useful to the creation of new sources and complex circuits for quantum technologies.
\end{abstract}

\maketitle

\section{Introduction}
\label{intro}
The development of new models and tools for machine learning (ML) is surprisingly affecting the study of many-body quantum systems and quantum optics~\cite{Preskill2021}. Neural networks (NN) enable representations of high-dimensional systems and furnish a universal ansatz for many purposes, like finding the ground state of many-body Hamiltonians~\cite{Carleo2019}, including dissipative systems~\cite{Vicentini2019,Mangini2021}.

Unsupervised and supervised learning endow new designs for quantum circuits~\cite{Marquardt2021}, metrology and cryptography~\cite{Sciarrino2018,Fratalocchi2021}, multilevel gates~\cite{Marcucci:20}, and Bell tests~\cite{Melnikov2020}.
NN are also triggering new fundamental investigations in quantum neuromorphic and wave computing~\cite{Marcucci2019b,Hughes2019,ballarini2019polaritonic,Nokkala2020,Markovic2020,Silva2021}, quantum thermodynamics~\cite{Sgroi2020}, and topological photonics~\cite{Pilozzi2021}.

The impact of ML in quantum optics and many-body physics is related to the versatile representation that the NN models furnish for functions of an arbitrary number of variables. Also, the powerful application programming interfaces (APIs), as {\tt TensorFlow}, enable many new features and tools to compute and design many-body Hamiltonians or large-scale quantum gates~\cite{Broughton2020}.

Here we show that NN models are also useful when considering representations in the phase space, as the characteristic functions $\chi$ or the Q-representation~\cite{BarnettBook}. Unitary operators, as squeezers or displacers, act on the phase-space as variable transformations that correspond to layers in the NN model. Hence, a multilayer NN may encode phase-space representations of complex many-body states. This encoding has two main advantages: on the one hand, one can quickly build complex quantum states by combining NN layers; on the other hand, one can use the automatic graph building and API differentiation technology to compute observables. Also, graphical and tensor processing units (GPU and TPU) may speed up the computation.

In the following, we show how to compute the probability of multi-particle patterns when Gaussian states propagate in a system made of squeezers and interferometers. This problem corresponds to the renowned Gaussian Boson sampling~\cite{Hamilton2016,Quesada2018}, which recently demonstrated the quantum advantage at an impressing scale~\cite{Zhong2020}, following earlier realizations~\cite{Tillmann2012,Broome2013,Spring2013,Spagnolo2014,Carolan2014,Wang2019} of the original proposal by Aharanson and Arkhipov~\cite{Aaronson2013}. The theory of Gaussian Boson sampling (GBS) heavily relies on phase-space methods~\cite{Kruse2018}, making it an exciting NN test-bed supported by recently reported trainable hardware~\cite{Arrazola2021,Hoch2021,Zhong2021}.

A notable outcome of adopting NN models in the phase space is the possibility of training multi-particle statistics~\cite{Arrazola2020} and other features as the degree of entanglement. Indeed, most of the reported investigations in quantum ML, focus either on using NN models as a variational ansatz or tailoring the input/output response of a quantum gate. On the contrary, ML in the phase space permits optimizing many-particle features, for example, to increase the probability of multi-photon events. NN may open new strategies to generate non-classical light or enhance the probability of observing large-scale entanglement with relevance in many applications. Here, we derive the NN representing the characteristic function of the Gaussian boson sampling setup. Proper NN training increases the photon-pair probability by orders of magnitude.

\begin{figure}
   \begin{center}
     \includegraphics[height=0.5\textwidth]{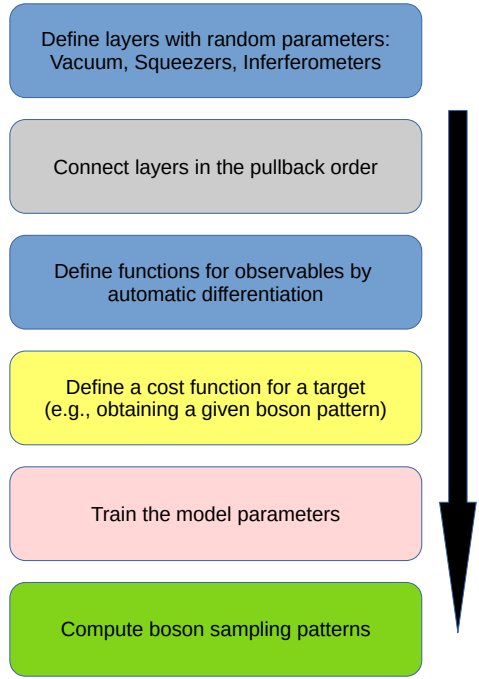}
     \end{center}
   \caption{Workflow~\label{fig:scheme} of the proposed
     methodology to train Boson sampling by representing the
   characteristic function as a neural network.}
 \end{figure}

 Fig.~\ref{fig:scheme} shows the general workflow of the proposed methodology,
 the different steps enable to define a trainable model for optmizing
 Gaussian boson sampling.
In Section~\ref{sec:chinn}, we introduce the way
we adopt a neural network to compute the characteristic function.
In Sec.~\ref{sec:obs}, we detail how to compute the
observable as derivatives of the characteristic function neural network.
In Sec.~\ref{sec:GBS}, we show how to compute the
Gaussian boson sampling patterns.
In Sec.~\ref{sec:trainGBS}, we introduce the loss function
and describe the training of the model to optimize specific patterns.
Conclusions are drawn in Sec.~\ref{sec:con}.
\section{Characteristic function as a neural network}\label{sec:chinn}
\begin{figure*}
  \begin{center}
    \includegraphics{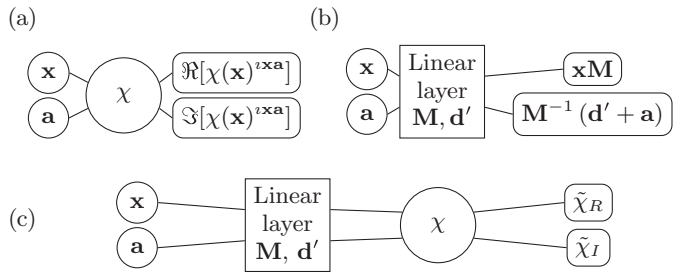}
    \end{center}
  \caption{(a) A neural network model for the characteristic function.
    Two inputs, a data vector $\mbfx$  with shape $1\times N$ and a bias vector $\mbfa$ with shape $N\times 1$ seed the model
    that compute $\chi$ and returns the real and imaginary parts of $\chi(\mbfx) e^{\imath \mbfx \mbfa}$.
    (b) A layer representing a linear transformation of the state by a unitary operator represented
    by a symplectic $N\times N$ matrix $\mbfM$ and a displacement $N\times 1$ vector $\mbfd'$. With such a definition layers can be cascaded, and one can represent single mode squeezers, interferometers, and other unitary operators.
    (c) A model representing a state with characteristic function $\chi$, subject to a unitary transformation. This is a pullback of a linear transform from the original state, which produces a new state with characteristic function $\tilde\chi$~[see Eq.~(\ref{eq:chitilde})].~\label{fig:1}}
\end{figure*}

In the phase space, we represent a $n$-body state by complex characteristic function $\chi(\mbfx)=\chi_R(\mbfx)+\imath \chi_I(\mbfx)$
of a real vector $\mbfx$~\cite{GardinerBook,BarnettBook}.
$\mbfx$ has dimension $1\times N$ with $N=2 n$.
For Gaussian states~\cite{X.2007}
\begin{equation}
  \chi(\mbfx)= e^{-\frac{1}{4}\mbfx\mbfg\mbfx^\top+\imath\, \mbfx\mbfd}.
\end{equation}
with $\mbfg$ the real covariance $N\times N$  matrix, and $\mbfd$ the  real displacement $N\times 1$ vector.
In our notation, we omit the symbols of the dot product such that $\mbfx\mbfd$ and $\mbfx\mbfg\mbfx^\top$ are scalars. One has ($j,k=0,1,2\dots,N-1$)
\begin{equation}
\langle \hat{R}_j\rangle =d_j= \left.\frac{\partial \chi}{\partial x_j}\right\vert_{\mbfx =0}\text{,}
\label{eq:chiIderivative}
\end{equation}
and
\begin{equation}
  g_{jk}=2\langle(\hat{R}_j-d_j)(\hat{R}_k-d_k)\rangle-\imath J_{jk},
  \end{equation}
  being ${\bf J}=\bigoplus_{j=0}^{n-1} {\bf J}_1$, ${\bf J}_1=\big(\begin{smallmatrix} 0 &1 \\ -1& 0\end{smallmatrix}\big)$~\cite{X.2007}. In Eq.~(\ref{eq:chiIderivative}), the canonical variables,  $\hat{q}_j=\hat{R}_{2j}$ and $\hat{p}_j=\hat{R}_{2j+1}$, with $j=0,1,\dots,n-1$, are organized in the $N\times 1$ operator array $\hat{\mbfR}$.
  As shown in Fig.~\ref{fig:1}a, the characteristic function is a NN layer with two real outputs $\chi_R$ and $\chi_I$. The $\chi$ layer has two inputs: $\mbfx$, and a auxiliary bias $N\times 1$ vector {\bf a}, for later  convenience. 

The vacuum state is a Gaussian state with $\mbfg={\bf 1}$ and $\mbfd={\bf 0}$.
From the vacuum, one can generate specific states by unitary operators,
as displacement or squeezing operators. These transform
the canonical variables as
$\hat{\widetilde{\bf R}}=\mbfM\, \hat{{\bf R}}+\mbfd'$,
where the symplectic matrix $\mbfM$ and the vector $\mbfd'$ depend on the specific operator (detailed, e.g., in \cite{X.2007}).
The characteristic function changes as
\begin{equation}
  \tilde\chi(\mbfx)=\chi(\mbfx\mbfM)e^{\imath \mbfx \mbfd'+\imath\mbfx \mbfa}=
  \chi(\mbfx\mbfM)e^{\imath (\mbfx\mbfM)\mbfM^{-1}(\mbfd'+\mbfa)}
  \label{eq:chitilde}
\end{equation}    
We represent the linear transformation as a NN layer with two inputs $\mbfx$ and $\mbfa$ and two outputs $\mbfx \mbfM$ and $\mbfM^{-1}(\mbfd'+\mbfa)$~(Fig.~\ref{fig:1}b). By this definition, Eq.~(\ref{eq:chitilde}) is as a two-layer NN.\@
Figure~\ref{fig:1}c shows $\tilde{\chi}$ as the ``pullback'' of the linear layer from the $\chi$ layer.
The two layers form a NN that can be implemented with common APIs.
\footnote{A {\tt TensorFlow} implementation in a {\tt Jupyter} notebook is available at \url{https://github.com/nonlinearxwaves/BosonSampling}}.
Given the vacuum state with characteristic function $\chi$, one can build the NN model of an arbitrary state by multiple pullbacks.
Indeed, we defined the linear layers in a way that they can be cascaded.
Figure~\ref{fig:2}a below shows a $n$-mode squeezed vacuum as a multiple pullback of single mode squezers, each acting on a different mode.
\section{Observables}\label{sec:obs}
Observables are computed as derivatives of the NN model. For example, the mean photon number per mode is related to the
derivatives of the characteristic function. The mean photon number for mode $j$, is
\begin{equation}
 \langle \hat{n}_j\rangle=-\left.\frac{1}{2}\left(\nabla_j^2+1\right)\chi\right|_{\mbfx=0}
  \end{equation}
  being $\nabla_j^2=\partial^2_{q_j}+\partial^2_{p_j}$
  and $q_j=x_{2j}$ and $p_j=x_{2j+1}$.
The differential photon number of modes $j$ and $k$ is
\begin{equation}
  \langle \left(\hat{n}_j-\hat{n}_k\right)^2\rangle=
\left.\left[\frac{1}{4}\left(\nabla_j^2-\nabla_k^2\right)^2-\frac{1}{2}\right]\chi\right|_{\mbfx=0}.
\end{equation}
Automatic differentiation packages enables an efficient computations of the derivatives of the NN model.
\section{Gaussian Boson sampling with the neural network model}\label{sec:GBS}
In the GBS protocol, one considers a many-body squeezed vacuum state propagating
in an Haar inteferometer, which distributes the photons in the output modes.
For modelling GBS, we hence need squeezing layers and a layer
representing the transmission through random interferometers.
The squeezing layers are realized by a proper design of the
corresponding symplectic matrices $\mbfM$ with $\mbfd=0$.
We implement the Haar matrix operator by QuTiP software~\cite{Johansson2013}. 
 Figure~\ref{fig:alg1} shows a pseudo-code to build the neural network model by composing different layer.
 \begin{figure}[ht!]
   \begin{algorithmic}
     \State Input = input layer\Comment{Create input layer}
     \State V = vacuum layer\Comment{Create vacuum layer}
     \State \For{i=0,1,2,\ldots,n-1}
     \State S[i] = squeezer layer for mode i\Comment{Create a squeezing layer per mode}
     \EndFor
     \State
     \State R = random interferometer\Comment{Create a random interferometer}
     \State x, a= Input\Comment{Define input tensors x and a}
     \State x, a= R (x,a) \Comment{Connect interferometer in pullback order}
     \State \For{i=0,1,2,\ldots,n-1} 
     \State x, a= S[i](x,a) \Comment{Connect squeezing layers in pullback order}
     \EndFor
     \State x, a= V(x,a) \Comment{Connect vacuum layer in pullback order}
   \end{algorithmic}
 \caption{Pseudo-code for the creation of a neural network representing a Gaussian boson sampling experiment\label{fig:alg1}}
 \end{figure}
\begin{figure*}
\includegraphics[width=0.75\textwidth]{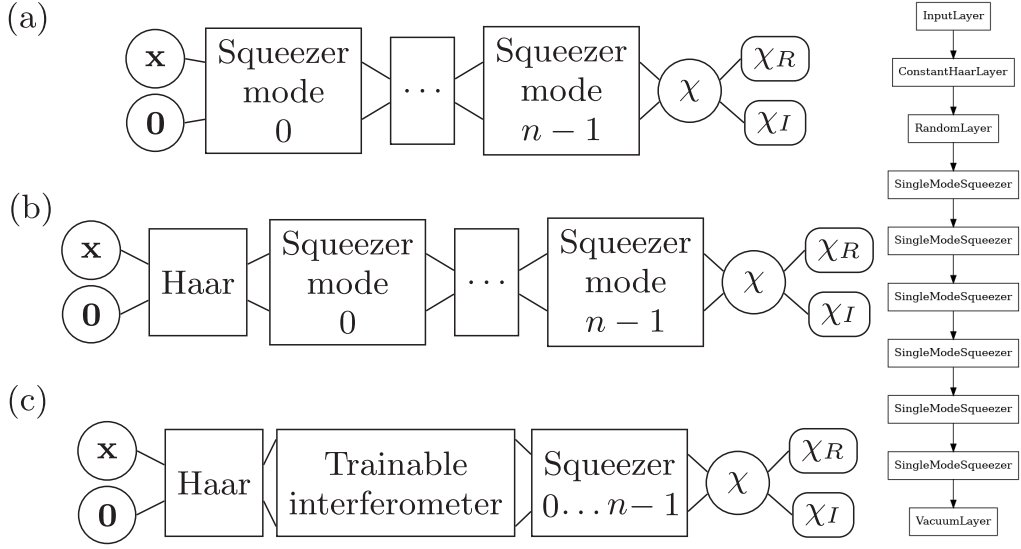}
\caption{(a) A multiple pullback that represents a many-body squeezed vacuum,
  obtained by a vacuum state $\chi$ by cascading $n$ identical single mode squeezers.
  The order of the squeezers is not relevant as they act of different modes.
  $\chi_R$ and $\chi_I$ are the real and imaginary part of the resulting characteristic
  function. (b) GBS setup, a $n$-body squeezed vacuum enters an Haar inteferometer. Note that the order of the operators, from the vacuum to the interferometer goes from right to left.
  (c) GBS setup including a trainable random interferometer before entering the Haar interferometer. The multiple squeezers are represented as a single block. The trainable interferometer can optimize the probability of pair generation.
  The right panel shows the architecture of the {\tt TensorFlow} model for $n=6$.~\label{fig:2}}
\end{figure*}

Fig.~\ref{fig:2}b is a graphical representation of the GBS NN model~\cite{Hamilton2016}.
Boson sampling corresponds to computing the probability $\Pr(\bar\mbfn)$ of finding $\bar{n}_0$ photons in mode $0$, $\bar{n}_1$ photons in mode $1$, and so forth. $\bar\mbfn=(\bar{n}_0,\bar{n}_1,\ldots,\bar{n}_{n-1})$ is a given photon pattern. Letting $\hat{\rho}$ the density matrix, one has \[\Pr(\bar\mbfn)=\Tr[\hat{\rho}|\bar\mbfn\rangle\langle\bar\mbfn|],\] with \[|\bar\mbfn\rangle\langle\bar\mbfn|=\otimes_{j=0}^{n-1}|\bar{n}_j\rangle\langle \bar{n}_j|.\]
Correspondingly~\cite{Kruse2018},
\begin{equation}
  \Pr(\bar\mbfn)=\left.\frac{1}{\bar{\mbfn}!}
\prod_{j=0}^{n-1}   {\left(\frac{\partial^2}{\partial \alpha_j\partial \alpha_j^*}\right)}^{\bar{n}_j}
    e^{\sum_j|\alpha|_j^2}Q_\rho(\mbfalpha,\mbfalpha^*)\right|_{\mbfalpha=0} 
\end{equation}
where $\bar\mbfn!=\bar{n_0}!\bar{n}_1!\ldots\bar{n}_{n-1}!$
and \[Q_\rho=\pi^n \langle \mbfalpha | \rho | \mbfalpha \rangle\]
is the Q-rapresentation of the density matrix~\cite{GardinerBook,BarnettBook} with $\mbfalpha=\left(\alpha_0,\alpha_1,\ldots,\alpha_{n-1}\right)$ complex displacements.

We introduce the $N\times 1$ real vector $\mbfk$ as
\[\begin{aligned}
    k_{2j} &= \frac{\alpha_j^*+\alpha_j}{\sqrt{2}}\\
  k_{2j+1} &= \frac{\alpha_j^*-\alpha_j^*}{\sqrt{2}\imath}
\end{aligned}\]
and we have
\begin{equation}
  \Pr(\bar\mbfn)=\left.\frac{1}{\bar{\mbfn}! 2^{\bar{n}_T}}
\left(\prod_j   {\tilde\nabla_j}^{2\bar{n}_j}\right)
e^{\frac{\mbfk^2}{2}}Q_\rho(\mbfk)\right|_{\mbfk=0}
\label{eq:Qrhox}
\end{equation}
with
$\tilde\nabla_j^2=\partial^2/\partial k_{2j}+\partial^2/\partial k_{2j+1}$
and $\bar{n}_{T}=\sum_{j=0}^{n-1}\bar{n}_j$.
$Q_\rho$ in Eq.~(\ref{eq:Qrhox}) can be evaluated explicitly
as a multidimensional Gaussian integral: 
\begin{equation}
  \Pr(\bar\mbfn) =
\left.  \frac{1}{\bar{\mbfn}! 2^{\bar{n}_T}}
  \left(\prod_j\tilde\nabla_j^{2\bar{n}_j}\right)
\mathcal{Q}(\mbfk)\right|_{\mbfk=0}
\label{eq:qtransform1}
  \end{equation}
with ($p,q=0,1,..,N-1$)
  \begin{equation}
    \mathcal{Q}(\mbfk)=
    \frac{1}{\sqrt{2^n \det A}}
    e^{\frac{1}{2}\mbfk^2}
    e^{-\frac{1}{2}\sum_{pq} A^{-1}_{pq} (k_p-d_q)(k_p-d_q)}
    \label{eq:qtransform2}
    \end{equation}
being $A_{pq}=\frac{1}{2}\left(g_{pq}+\delta_{pq}\right)$.  
Eq.~(\ref{eq:qtransform1}) and (\ref{eq:qtransform2}) can be implemented
as further layers of the NN, and the probability of a given pattern
computed by running the model.
Figure~\ref{fig:3}a shows an example of the pattern probability distribution
with $n=6$, obtained by using the NN model in Fig.~\ref{fig:2}b with
squeezing parameters $r_j=0.88$ and $\phi_j=\pi/4$, such that
all the single mode squeezers are identical, each with mean photon number $\sinh(r_j)^2\simeq 1$.
As in \cite{Hamilton2016}, we consider patterns with $\langle \hat{n}_j\rangle=\{0,1\}$.
\section{Training Gaussian boson sampling}
\label{sec:trainGBS}
Our interest is understanding if we can train the model to maximize
the generation of specific patterns, e.g., a photon pair in modes $0$ and $1$.
Using complex media to tailor linear systems is a well renowned technique as, for example,
to synthesize specific gates~\cite{Gigan2019,Taballione:19} or taming entanglement~\cite{Valencia2019}.
Here, we use the NN model in the phase space to optimize multi-particle events.

One could use the squeezing parameters in the model in Fig.~\ref{fig:2}b as
training parameters. However, the degree of squeezing
affects the number of particles per mode, and we want to alter
the statistical properties of states without changing the average number
of particles. We hence consider a GBS setup with an additional trainable
interferometer as in Fig.~\ref{fig:2}c, which is typically realized by
amplitude or phase modulators.

In Fig.~\ref{fig:2}c, $n$ squeezed vacuum modes impinge
on a trainable interferometer and then travel through a Haar interferometer. Instead of two distinct interferometers, one could use a single device (i.e., combine the Haar interferometer with the trainable interferometer), but we prefer to distinguish the trainable part from the mode-mixing Haar unitary operator.

Given $n$ modes, our goal is to maximize the probability of patterns that
contains a pair of photons in the mode $0$ or $1$.
For example, for $n=6$, this means maximizing the probability of $\bar\mbfn=(1,1,0,0,0,0)$ with respect to
$\bar\mbfn=(1,0,0,1,0,0)$. We use as loss function
\begin{equation}
  \mathcal{L}=e^{\langle(\hat{n}_0-\hat{n}_1)^2\rangle}
  \label{eq:1}
\end{equation}
which is minimal when the expected differential number of photons
in mode $0$ and mode $1$ vanishes. This is the case
when the state has a particle pair in mode $0$ and mode $1$.
We stress the difference in using other cost functions,
which involve the expected number of photons per mode as, e.g.,
\[
  \mathcal{L}_0=e^{{(\langle \hat{n}_0\rangle-\langle \hat{n}_1\rangle)}^2}.
  \label{eq:1b}
  \]
The linear interferometer does not affect the average
number of photons (which are mixed by the Haar layer).
Correspondingly, training using $\mathcal{L}_0$ Eq.~\ref{eq:1b} is not be effective to generate entangled pairs. On the contrary, $\mathcal{L}$ in Eq.~(\ref{eq:1}) contains $\langle \hat{n}_0\hat{n}_1\rangle$, which is maximal with a photon pair in modes $0$ and $1$.

\begin{figure*}
  \begin{center}
    \includegraphics[width=0.6\textwidth]{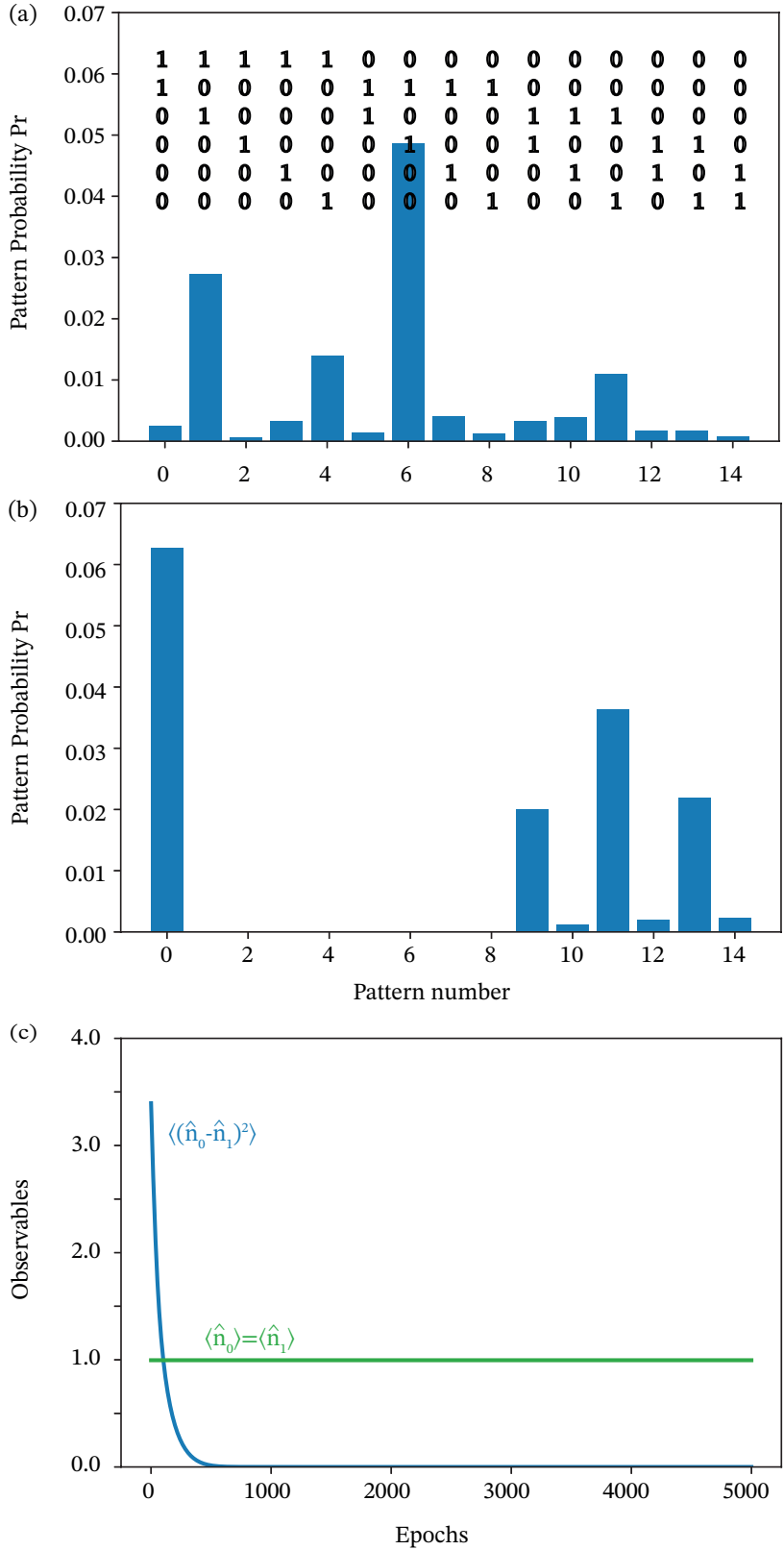}
    \end{center}
  \caption{(a) Probability distribution of patterns with two photons for $n=6$ in the model in Fig.~\ref{fig:2}c, before training.
    The insets detail the particle distribution in the patterns.
  (b) As in (a) after training, the probability of finding a pair in mode $0$ and $1$ is enhanced
  by more than one order of magnitude.
  (c) Mean photon number in mode $0$ and $1$ during the training epochs (green), and
  expected differential photon number $\langle \left(\hat{n}_0-\hat{n}_1\right)^2\rangle$ in the two modes,
  which vanishes after thousands of epochs. The statistical distribution of pairs changes at a constant photon number per mode.
  Data generated by the code in \url{https://github.com/nonlinearxwaves/BosonSampling}.}\label{fig:3}
\end{figure*}

Fig.~\ref{fig:3}a shows the computed probabilities of pairs for the model in Fig.~\ref{fig:2}c, with a random instance of the Haar and the linear inteferometers. Training strongly alters this statistical distribution, as shown in Fig.~\ref{fig:3}b.
Fig.~\ref{fig:3}c shows the trend during
the training epochs of $\langle(\hat{n}_0^2-\hat{n}_1^2)\rangle$,
which goes to zero while the mean photon numbers $\langle\hat{n}_0\rangle$ and
$\langle\hat{n}_1\rangle$ remain unaltered.

\begin{figure*}
  \begin{center}
    \includegraphics[width=0.5\textwidth]{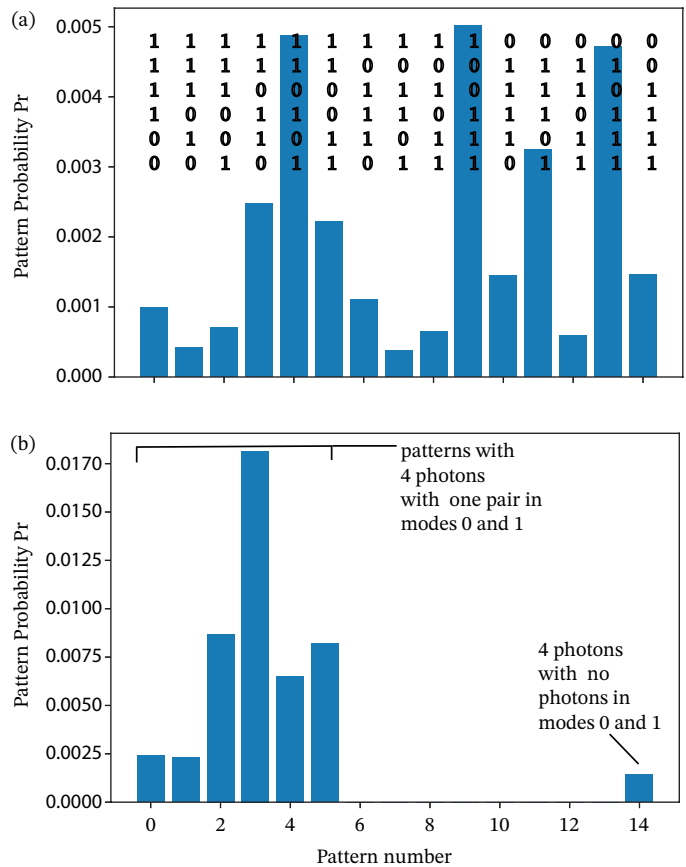}
    \end{center}
  \caption{(a) Probability distribution of patterns with $4$ photons
    ($n=6$) in the model in Fig.~\ref{fig:2}c before training.      
      The insets detail the particles in each pattern.
      (b) As in (a) after training; the probability of patterns
      with two photons in modes $0$ and $1$ is maximized.
      Data generated by the code in \url{https://github.com/nonlinearxwaves/BosonSampling}.}\label{fig:BS7quater}
\end{figure*}

Training also maximizes higher photon events,
as in the pattern $\bar\mbfn=(1,1,1,1,0,0)$ with $4$ photons and $n=6$. Fig.~\ref{fig:BS7quater}a shows the pattern probability with $4$ photons.
After training with the loss function in Eq.~(\ref{eq:1}), $\Pr(\bar\mbfn)$ substantially increases for the patterns with
four photons containing $1$ pair in modes $0$ and $1$~(Fig.~\ref{fig:BS7quater}b).
\section{Conclusions}\label{sec:con}
We have shown that a many-body characteristic function may be reformulated as a layered neural network.
This approach enables to build complex states for various applications, as gate design or boson sampling.

A common argument in criticizing quantum neural networks is that the linear quantum mechanics does not match with the nonlinearity-eager NN models.
However, recent investigations show that nonlinearity may be introduced
in quantum neural networks~\cite{Zhao2021}.
Our remark is that if we formulate quantum mechanics in the phase space, nonlinearity arises in the characteristic function (or other representation). We analyzed this strategy in the simplest case of Gaussian states. The resulting model is universal and may be trained for different purposes.
For this reason, phase space models allow naturally in dealing with non-classical states and computing observables by derivatives. This formulation opens many opportunities. For example, the optimization of multi-particle events can be extended to fermionic fields.
As a drawback, computing boson patterns probabilities by NN APIs is not expected to be competitive with highly optimized algorithms running on large-scale clusters~\cite{Quesada2020,Li2020}. Still, it appears to be a versatile and straightforward methodology. 

Here, we have shown many-body quantum state design and engineering by {\tt TensorFlow}. We have demonstrated how to enhance multi-particle generation, with many potential applications in quantum technologies. In addition, the proposed method enables training Boson sampling without explicitly computing derivatives of the Hafnian~\cite{Arrazola2020,Broughton2020}, but resorting to automatic computational packages. We have tested the algorithm with a conventional
workstation with a single commercial GPU (NVIDIA QUADRO RTX 4000),
with a computational time of the order of few minutes with 6 modes.

The method can be generalized to other boson sampling setups, as including
Glauber layers and multi-mode squeezers. Also, it readily allows to test
different loss functions for tailoring the boson sampling patterns.
Extension beyond Gaussian states can be envisaged by using a general machine learning networks with an arbitrary number of layers and different nonlinearity.  

\begin{acknowledgements}
We acknowledge support from Horizon 2020 Framework Programme QuantERA grant QUOMPLEX, by National Research Council (CNR), Grant 731473.
\end{acknowledgements}

%
%

\bibliographystyle{spphys}       

\end{document}